\documentclass[sigconf,authordraft,review=false]{acmart}
\usepackage{algorithmic}
\usepackage{algorithm}
\usepackage{graphicx}
\usepackage{textcomp}
\usepackage{xcolor}
\usepackage[symbol]{footmisc}
\AtBeginDocument{%
  \providecommand\BibTeX{{%
    \normalfont B\kern-0.5em{\scshape i\kern-0.25em b}\kern-0.8em\TeX}}}

\setcopyright{acmcopyright}
\copyrightyear{2018}
\acmYear{2018}
\acmDOI{10.1145/1122445.1122456}

\acmConference[Woodstock '18]{Woodstock '18: ACM Symposium on Neural
  Gaze Detection}{June 03--05, 2018}{Woodstock, NY}
\acmBooktitle{Woodstock '18: ACM Symposium on Neural Gaze Detection,
  June 03--05, 2018, Woodstock, NY}
\acmPrice{15.00}
\acmISBN{978-1-4503-XXXX-X/18/06}

\begin{document}

\title{IR-BERT: Leveraging BERT for Semantic Search in Background Linking for News Articles}

\author{Anup Anand Deshmukh}
\email{aa2deshmukh@uwaterloo.ca}
\affiliation{%
  \institution{University of Waterloo}
  \streetaddress{David R. Cheriton School of Computer Science}
  \city{Waterloo}
  \state{Ontario}
  \postcode{N2L3K3}
   \country{Canada}
}

\author{Udhav Sethi}
\email{udhav.sethi@uwaterloo.ca}
\affiliation{%
  \institution{University of Waterloo}
  \streetaddress{David R. Cheriton School of Computer Science}
  \city{Waterloo}
  \state{Ontario}
  \postcode{N2L3K3}
  \country{Canada}
}


\begin{abstract}
This work describes our two approaches for the background linking task of TREC 2020 News Track. The main objective of this task is to recommend a list of relevant articles that the reader should refer to in order to understand the context and gain background information of the query article. Our first approach focuses on building an effective search query by combining weighted keywords extracted from the query document and uses BM25 \cite{robertson2009probabilistic} for retrieval. The second approach leverages the capability of SBERT \cite{reimers2019sentence} to learn contextual representations of the query in order to perform semantic search over the corpus. We empirically show that employing a language model benefits our approach in understanding the context as well as the background of the query article. The proposed approaches are evaluated on the TREC 2018 Washington Post dataset and our best model outperforms the TREC median as well as the highest scoring model of 2018 in terms of the nDCG@5 metric. We further propose a diversity measure to evaluate the effectiveness of the various approaches in retrieving a diverse set of documents. This would potentially motivate researchers to work on introducing diversity in their recommended list. We have open sourced our implementation on Github \footnote{\url{https://github.com/Anup-Deshmukh/TREC_background_linking}} and plan to submit our runs for the background linking task in TREC 2020.

\end{abstract}



\maketitle

\section{Introduction}
In last few years, online news services have been key sources of information and have affected the way we receive and share news. While drafting a news article, often times it is assumed that reader has a background knowledge about the article's story. This demands the need of linking other articles that set the background context for the article in focus. These articles are typically by the same author, provide extra relevant information or introduce us to the key ideas to help understand the topic better. However, defining what could provide "background context" and retrieving such documents is not straightforward. 

Motivated by this need and with the goal of fulfilling the search requirement of news readers, the Background Linking task was introduced in the NEWS track of TREC 2018. The main objective of this task is to recommend/retrieve a list of other articles that the reader should refer to in order to understand the context and gain background information of the query article.

In this paper, we discuss our proposed methods to tackle the problem of background linking. The paper is structured as follows: In section 2, we provide an overview of the earlier submissions and some other noteworthy methods that motivate our idea. In section 3, we demonstrate in detail our retrieval strategies followed by section 4 and 5, where we report the retrieval performances and show their effectiveness compared to earlier submissions. In the end we summarize and conclude our work in section 6. 

\section{Related Work}

BM25 \cite{robertson2009probabilistic} is one of the most popular ranking functions used by search engines to estimate the relevance of documents to a given search query. It is based on a bag-of-words retrieval function that ranks a set of documents based on the query terms appearing in each document, regardless of their proximity within the document. A number of previous approaches to background linking are built using BM25. Anserini \cite{yang2017anserini, anserini2} is an open-source information retrieval toolkit built on Lucene. \cite{yang2018anserini} uses Anserini to solve the background linking problem, by exploring different approaches to constructing a keyword query from a query article, and using the BM25 ranking function to retrieve the documents relevant to the constructed query. ICTNET \cite{DBLP:conf/trec/DingLZLDH19} and htwsaar1 \cite{bimantara2018htw} follow a \textit{tf-idf} based approach similar to Anserini to form its query while bigIR \cite{essambigir} bases the construction of the search query mainly on a graph-based analysis of the query article’s text. Other approaches using BM25 include UDInfolab\_ent \cite{lualeveraging} and DMINR \cite{missaouidminr}, which focus on leveraging Named Entities (NEs) in the query article to form a query before using BM25 to identify the background articles. The best performing model for TREC 2018 is \textit{umass\_cbrdm}, which is an RM model \cite{lavrenko2017relevance} with BM25 scoring functions.
\begin{figure*}[h]
  \label{fig:preprocess}
  \centering
  \includegraphics[width=0.8\textwidth]{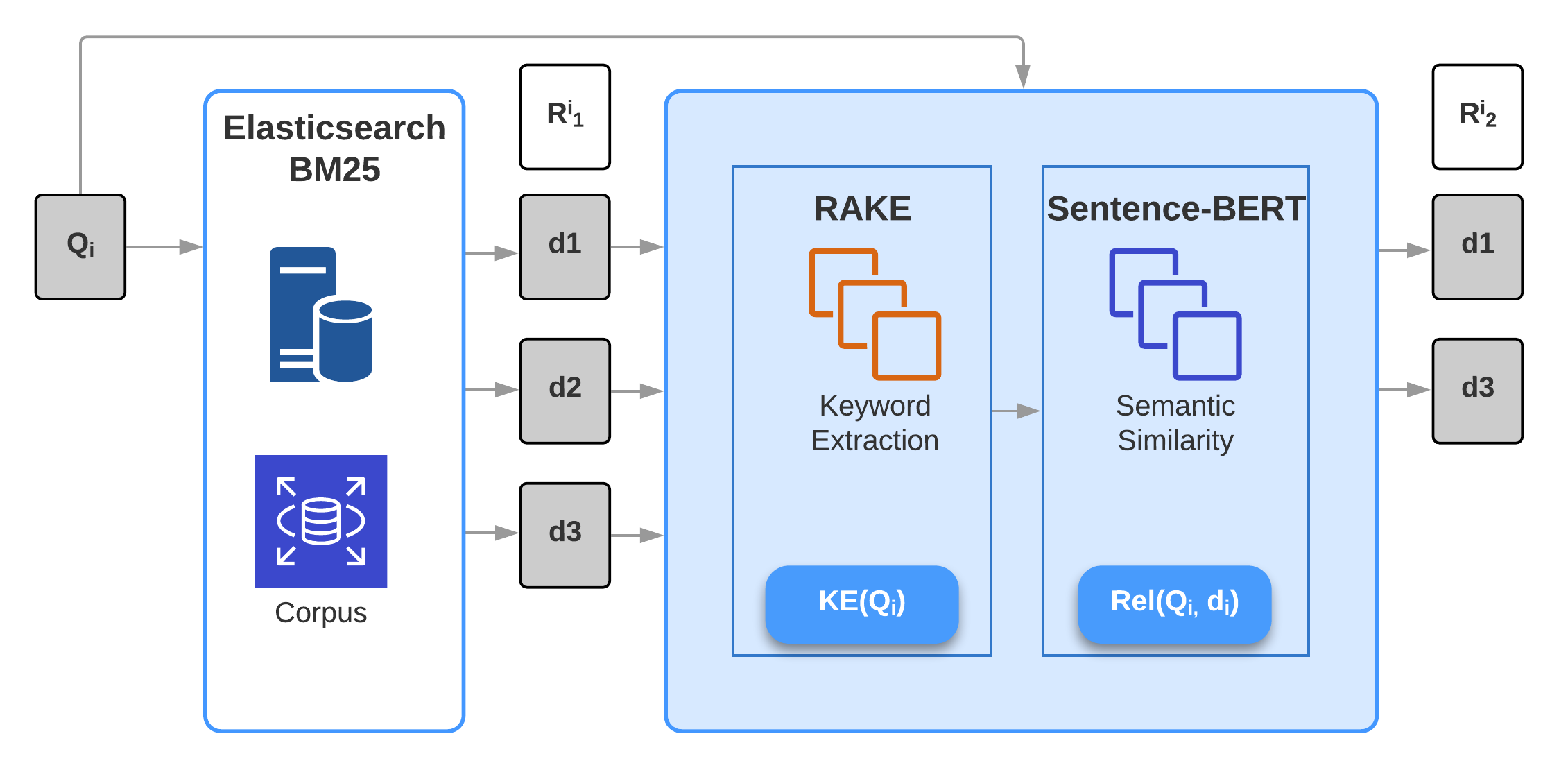}
  \caption{IR-BERT pipeline \label{fig:final}}
\end{figure*}

There also exist methods which have tried exploiting language models for the task of ad-hoc retrieval \cite{yang2018anserini, nogueira2019multi}. BERT \cite{devlin2018bert} is a state of the art language model which has achieved groundbreaking results in many NLG and NLI tasks. BERT has been further fine tuned for downstream tasks and used in many applications. One such relevant application is finding the semantic similarity between text documents since both query and results to be retrieved are textual within the problem of background linking. Sentence-BERT \cite{reimers2019sentence} is one such model which enable us to derive the semantically meaningful sentence embeddings by modifying original BERT using Siamese networks. With Sentence-BERT we can now take advantage of BERT embeddings for the tasks like semantic similarity comparison and information retrieval via semantic search.

\section{Methodology}

The background linking task can be formulated as the following: \textit{Given a news story $S$ and a collection of news articles $A$, retrieve other news articles from $A$ that provide important context or background information about $S$.}

It is reasonable to consider the background linking task as a specific case of news recommendation task aimed at retrieving relevant articles from a corpus $A$ for queries generated from an article $S$. For this task, in order to retrieve articles that can provide contextual information, we filter out forward links from our results, i.e., the articles published after the query article are not considered.

There are two main components to our solution for this task of background linking \textit{a)} Constructing a search query $Q$ from the document $S$ and \textit{b)} Performing a search in $A$ against $Q$ to retrieve articles providing background information on $S$. In sections \ref{sec:app1} and \ref{sec:app2}, we describe two approaches that focus on these two steps separately and attempt to show improvement in order to solve the background linking task.

\subsection{Approach 1 (A1): Weighted Search Query + BM25}
\label{sec:app1}
Our first approach is based on BM25, similar to \cite{yang2018anserini}, where we focus on building an effective search query that best captures the relevant topics of the query article. The problem is formulated as extracting the essential keywords from the query article, assigning them weights according to their relevance, and concatenating them to form a query. This query is then used to search the corpus using BM25 through which a final ranked list is generated.

\paragraph{Query Construction}
To find the keywords $\{k_1, ... k_n\}$ from query document $S$, we sort all the words in $S$ in decreasing order of their \textit{tf-idf} score. To assign different relevance levels to each keyword, we define a weight $w_i$ for each keyword $k_i$ as follows:
\begin{equation} \label{eqn:1}
  w_i =  nint \left( \frac{s_i}{\sum_{j=1}^{n} s_j} \cdot n \right)
\end{equation}
\begin{equation} \label{eqn:2}
    s_i = tf(k_i, S) \cdot idf(k_i, A)
\end{equation}

where $n$ is the number of keywords, and $tf$ and $idf$ are the two statistics, term frequency and inverse document frequency, respectively. To apply the weight for each keyword, we round its value to the nearest integer $w_i$ and repeat the $i$th keyword $k_i$, $w_i$ number of times in the query. This weighted query is fed to BM25 to retrieve the top $t$ number of articles. We experiment with different lower and upper bounds on the number of repetitions as discussed in section \ref{sec:expt-tune} and finally set the lower bound as 1 and the upper bound as 5.

We also assign different weights to the contribution of the title and body of the article $S$ in query $Q$. Our experiments with different values for these weights are discussed in section \ref{sec:expt-tune}. We achieve the best scores by setting the weight of the title and 0.7 and the body as 0.3.

\subsection{Approach 2 (A2): IR-BERT}
\label{sec:app2}
Approach 1 is entirely based on the term frequencies of words appearing in the query article where BM25 then simply does search and retrieval in the indexed corpus. In order to understand the context of the query article $S$, it is important to take semantics of words into consideration. This is because the background articles may not necessarily contain the keywords in search query $Q$ constructed from the query article $S$. For example, a query article which goes by the title \textit{In Russia, political engagement is blossoming online} is likely to have \textbf{Russia} and \textbf{online} in the constructed query. In order to find the background articles, the retrieval model first must understand that \textbf{Russia} is a \textit{country} and \textbf{online} refers to \textit{social media} platforms like \textit{Facebook}, \textit{Twitter} etc. which are based on \textit{internet}.

To this end, we propose IR-BERT, which combines the retrieval power of BM25 with the contextual understanding gained through a BERT based model. Similar to Approach 1, using weighted query $Q_i$ we first retrieve $p$ number of documents using BM25. Let this set of documents be $R^i_1 = \{d_1, d_2, .., d_p\}$ where $|R^i_1| = p$. The goal now is to carry out the semantic search of $Q_i$ over the set of retrieved documents $R^i_1$ to arrive at the final set of documents $R^i_2 = \{d_1, d_2, .., d_t\}$ where $|R^i_2| = t$. The overall pipeline of IR-BERT is shown in Figure \ref{fig:final}. 

\paragraph{RAKE: } Before carrying out the semantic search over the set of documents $R^i_1$, it is important to feed only those words to sentence-BERT whose semantic meaning could benefit us. Thus, every document in $R^i_1$ is passed through the Rapid automatic Keyword Extraction (RAKE) algorithm \cite{rose2010automatic}. RAKE is an algorithm which takes a list of stopwords and the query as inputs and extracts keywords from the query in a single pass. The reason we chose to use RAKE is that it is completely domain independent. It is based on the idea that co-occurrences of words are meaningful in determining whether they are keywords or not. The relations between the words are hence measured in a manner that automatically adapts to the style and content of the text. This allows RAKE to have adaptive measurement of word co-occurrences which are used to score candidate keywords. 

\begin{figure}[h]
  \centering  
  \includegraphics[width=0.8\linewidth]{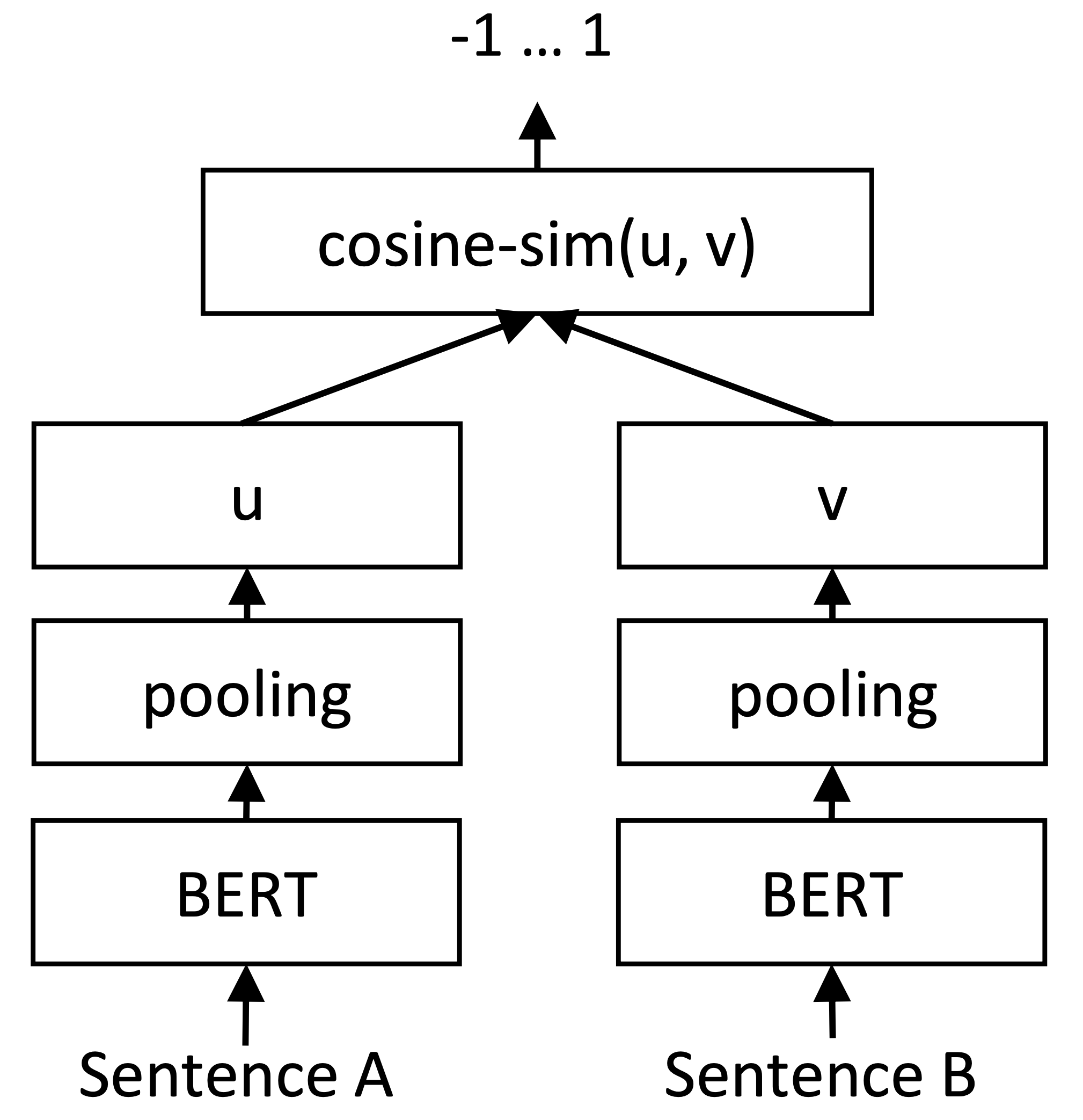}
  \caption{Sentence-BERT \cite{reimers2019sentence} architecture at Inference (to calculate the similarity scores) \label{fig:sen}}
\end{figure}

\paragraph{Sentence-BERT (SBERT): } Sentence-BERT \cite{reimers2019sentence} adds a pooling operation on top of the last layer of BERT and is fine tuned to derive a fixed size sentence embedding. \cite{reimers2019sentence} further uses Siamese and triplet networks \cite{schroff2015facenet} to update the weights of this model. Authors of Sentence-BERT incorporate the Siamese networks in order to have sentence embeddings that are semantically meaningful and which in turn can be compared with cosine-similarity. Figure \ref{fig:sen} describes the Sentence-BERT architecture.

The updated retrieved documents in $R^i_1$ given out by RAKE only has keywords and this set of keyword for every document can be treated as a sentence. Thus SBERT is used to obtain the embeddings of $R^i_1$ and $Q_i$. The documents in $R^i_1$ are then sorted according to their cosine similarity with the query $Q_i$ (equation \ref{eqn:cos}). We further normalize this similarity measure using equation \ref{eqn:score}. Generating the final list of documents $R^i_2$ via SBERT embeddings consist of steps mentioned in algorithm \ref{code:one}. 

\begin{equation} \label{eqn:cos}
  CosineSim(e_1, e_2) = \frac{e_1.e_2}{||{e_1}||.{||e_2||}}
\end{equation}

\begin{equation} \label{eqn:score}
  Score = \frac{1}{1+e^{-100 (CosineSim(e_1, e_2) - 0.95)}}
\end{equation}

\begin{algorithm}
\begin{algorithmic}[1]
\caption{ComputeRelevance($Q_i$, $R^i_1$ ) \label{code:one}}
\label{clus_algo}
    \STATE $p \gets$ Number of documents retrieved by BM25 
    \STATE $t \gets$ Required number of final documents  
    \STATE $q_i \gets$ SBERT($Q_i$)
    \FOR{i=1, \ldots p}{
        \STATE $d_k$ = $R^i_1[i]$ 
        \STATE $E_i$ = SBERT$(d_k)$
        \STATE $f_i$ = Score($E_i, q_i$)
        \STATE $i \gets i+1$
        }
    \ENDFOR
    \STATE $F \gets$ sorted list of $f_i$
    \STATE $R^i_2$ gets top $t$ documents with corresponding indices in $F$ 
    \RETURN $R^i_2$
\end{algorithmic}
\end{algorithm}

\section{Experimental Setup}

 \begin{figure*}[h]
  \centering
  \includegraphics[width=0.7\textwidth]{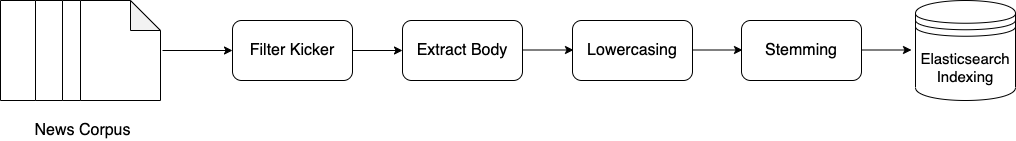}
  \caption{Data Preprocessing Steps \label{fig:preprocess}}
\end{figure*}

\subsection{Dataset Preprocessing}
We used the Washington Post Corpus\footnote{\url{https://trec.nist.gov/data/wapost/}} released by TREC for the 2018 news track for our experiments and preprocessed it using the steps shown in Figure \ref{fig:preprocess}. The collection contains 608,180 news articles and blog posts from January 2012 through August 2017. The articles are in JSON format, and include fields for title, date of publication, kicker, article text, and links to embedded images and multimedia. All of our methods rely on Elasticsearch\footnote{\url{https://www.elastic.co/}} as the indexing platform. During indexing, we extracted the information from the various fields and indexed them as separate Elasticsearch fields. We also created a new field to store the body of the article. For this, we first extracted the HTML text content from the fields marked by type 'sanitized\_html' and subtype 'paragraph', and then concatenated them after using regular expressions to extract the raw text from HTML text. Next, we performed lower-casing, stop-word removal, and stemming on the raw text. The final preprocessed text was then indexed as a separate text field in Elasticsearch, representing the article body.

While indexing, we filtered out the articles from the "Opinion", "Letters to the Editor", or "The Post's View" sections, as labeled in the "kicker" field, are they are stated as not relevant in the TREC guideline. We used the the default scoring method in Elasticsearch, BM25, as the retrieval model. Also, Elasticsearch boosting queries were used to assign weights for title and body in the search queries.

\subsection{Evaluation Metrics}
The primary metric used by TREC for the background linking task is nDCG@5, where the gain value is
$2^{r-1}$ where r is the relevance level, ranging from $0$ (The linked document provides little or no useful background information) to 4 (The document MUST appear in the sidebar otherwise critical context is missing). The zero relevance level contributes no gain. Apart from this, we also report nDCG@10, Precision@5 and Precision@10 for some of our experiments.

\subsection{Proposed Diversity Measure}

As per TREC guidelines, one of the criteria for ranking for the background linking task is to have a retrieved list of articles which are diverse. The idea of diversity may seem subjective but it is possible to formulate the diversity measure. We define diversity by equation \ref{eqn:3}. 

\begin{equation} \label{eqn:3}
   Diversity = \frac{1}{|Q|}\sum_{Q_i}\frac{1}{|R^i_2|}\sum_{i\in R^i_2}\sum_{j\in {R^i_2,{j \neq i}}} distTFIDF(i, j)
\end{equation}

where $Q$ is the set of all queries/topics in TREC 2018. For every retrieved document list $R^i_2$ we calculate the sum of distance between all the possible pairs of documents in $R^i_2$. This is then summed over all queries/topics.

\subsection{Parameter Tuning}
\label{sec:expt-tune}
For both our proposed approaches in this work, we tuned a number of parameters using the TREC 2018 topics and relevance judgements. In this section, we list out the different settings under which we tested the two approaches. The P@5 and nDCG@5 results for our fine-tuning experiments are shown in Section \ref{sec:appendix}.

\paragraph{Parameters for Approach 1 (A1):}
\begin{itemize}
    \item Number of words in the query constructed from given news article.
    \item Relative weights of title and body in the query.
    \item Maximum and minimum number of repetitions of extracted keywords.
\end{itemize}

\paragraph{Parameters for Approach 2 (A2):} \begin{itemize}
    \item Number of filtered results from BM25.
    \item Minimum number of keywords generated from RAKE
\end{itemize}

\section{Results}

In our first set of experiments we compare our best performing models from two approaches, A1 and A2, with several other previous methods. First entry in Table \ref{tab:res1} represents the TREC 2018 hypothetical run that achieved a median effectiveness over all queries. The second group of entries represent some of the results for official runs in the TREC 2018 news track. Second row corresponds to the run where entire query document (without assigned weights) is directly fed to BM25 for retrieval. anserini\_1000w  \cite{yang2017anserini} is the best performing run submitted by the researchers at University of Waterloo, which also uses BM25 as the ranking function. umass\_cbrdm \cite{lavrenko2017relevance} represents the best performing model for the news track in TREC 2018 . While A1.1 uses only weighted body and title while constructing a query for BM25, A1.2 uses also uses weights for all the words present in the query document. A2.1 and A2.2 are models based on the Approach 2 and use RoBERTa \cite{liu2019roberta} and BERT sentence level embeddings respectively.

\begin{table}[h]
  \caption{Comparison of nDCG@5 between proposed methods and previous submissions \label{tab:res1}}
  \label{tab:ndcg}
  \begin{tabular}{ccccl}
    \toprule
    Methods & nDCG@5 \\
    \midrule
    \texttt{{TREC 2018 Median}} & 0.3448 \\
    \hline 
    \texttt{{BM25}} & 0.3251 \\
    \texttt{{anserini\_1000w}} & 0.3529 \\
    \texttt{{umass\_cbrdm}} & 0.4173 \\
    \hline 
    \texttt{{(A1.1) wBT+BM25}} & 0.4088 \\
    \texttt{{(A1.2) wQ+BM25}} & 0.3942  \\
    \texttt{{(A2.1) IR-RoBERTa}} & 0.394 \\
     \texttt{{(A2.2) IR-BERT}} & \textbf{0.4199}  \\
    \bottomrule
  \end{tabular}
\end{table}

A1.1, A1.2 and A2.1 significantly outperform anserini\_1000w and BM25, showing that our approach helps construct effective queries which represent the important topics in the query document. Furthermore, our best performing model A2.2 (IR-BERT) outperforms umass\_cbrdm, setting the new benchmark on 2018 background linking qrels. Also, all our runs outperform the TREC 2018 median for this task.

\begin{table}[h]
  \caption{Comparison of nDCG and precision values between proposed methods and BM25 \label{tab:res2}}
  \begin{tabular}{ccccl}
    \toprule
    Methods & nDCG@5 & nDCG@10 & P@5 & P@10 \\
    \midrule
    \texttt{{BM25}} & 0.3251 & 0.3359 &  0.532 & 0.446 \\
    \hline
    \texttt{{(A1.1) wBT+BM25}} & 0.4088 & 0.4155 &  \textbf{0.644} & 0.53 \\
    \texttt{{(A1.2) wQ+BM25}} & 0.3942 & \textbf{0.4315} &  0.632 & \textbf{0.578} \\
    \texttt{{(A2.1) IR-RoBERTa}} & 0.394 & 0.3918 &  0.628 & 0.514 \\
    \texttt{{(A2.2) IR-BERT}} & \textbf{0.4199} & 0.4104 &  0.628 & 0.504 \\
    \bottomrule
  \end{tabular}
\end{table}

In our next set of experiments, we list nDCG@5, nDCG@10, Precision@5 and Precision10 scores achieved by our approaches on the 2018 Washington Post dataset in table \ref{tab:res2}. A1.1 (wBT+BM25) achieves the best P@5 score. On the other hand, A1.2 (wQ+BM25) gives best performance over two other metrics. It is interesting to note that using RoBERTa for finding semantic similarity between query and documents harms the performance given by A1.1 and A1.2.

\begin{table}[h]
  \caption{Comparison of diversity of retrieved documents from proposed methods and BM25 \label{tab:res3}}
  
  \begin{tabular}{ccccccl}
    \toprule
    Methods &  \texttt{{BM25}} & \texttt{{wBT+BM25}}   & \texttt{{wQ+BM25}} & \texttt{{IR-RoBERTa}} & \texttt{{IR-BERT}} \\
    \midrule
    Diversity & 0.907  & 0.9067  & 0.912  &  \textbf{0.921}  & 0.9084  \\
    \bottomrule
  \end{tabular}
\end{table}

In our last set of experiments we evaluate the diversity of all our models along with that of BM25. It can be observed that A2.1 (IR-RoBERTa) which relatively fails on traditional measures, retrieves the most diverse list of background articles for a given query.

\section{Conclusion}
In this paper, we described two methods to solve the background linking task in the context of participating in the TREC 2020 news track. While our first approach attempts to extract representative keywords from the query article and use them to retrieve the article's background links, the second approach leverages the contextual understanding ability of BERT to perform semantic search for background articles on the corpus. Our best model, IR-BERT, achieved an nDCG@5 score of $0.4199$ beating the TREC median as well as the best performing model of 2018 on the TREC Washington Post 2018 dataset. We further measured the diversity of the retrieved background articles for our approaches using a diversity measure based on \textit{tf-idf}.

\bibliographystyle{ACM-Reference-Format}
\bibliography{main_file}

\newpage

\section{Appendix}
\label{sec:appendix}

\begin{figure}[h]
  \centering  
  \includegraphics[width=\linewidth]{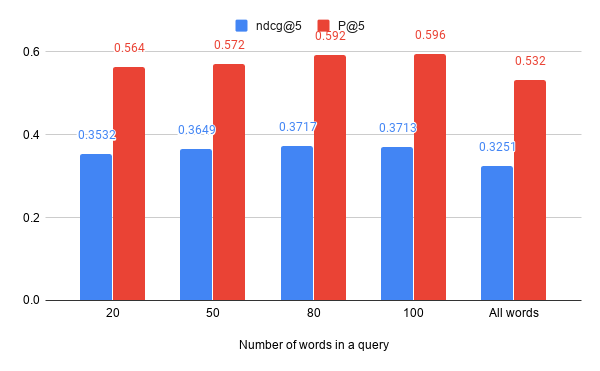}
  \caption{Fine tuning number of words in a query for BM25 \label{fig:ap1}}
\end{figure}

\begin{figure}[h]
  \centering  
  \includegraphics[width=\linewidth]{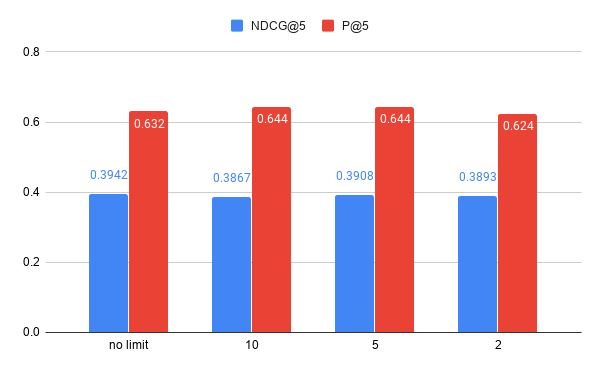}
  \caption{Fine tuning the maximum weight (repetitions) for a term in the query in Approach 1 \label{fig:ap1}}
\end{figure}

\begin{figure}[h]
  \centering  
  \includegraphics[width=\linewidth]{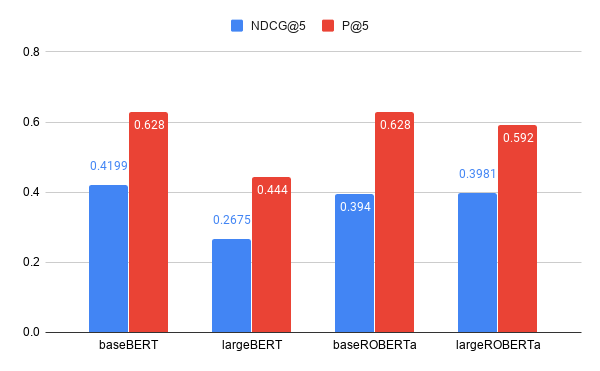}
  \caption{Results with different BERT models in Approach 2 \label{fig:ap1}}
\end{figure}

\end{document}